\documentclass[prl,twocolumn,aps,floatfix,longbibliography,nofootinbib,superscriptaddress]{revtex4-1}
\usepackage{pslatex,graphicx,dcolumn,bm,natbib,amssymb,amsmath,color,mathtools}
\usepackage[utf8]{inputenc}
\usepackage[T1]{fontenc}
\usepackage{soul}
\usepackage{comment}
\usepackage{hyperref}
\newcommand{\be}{\begin{equation}}
\newcommand{\ee}{\end{equation}}

\usepackage{calrsfs} 
\DeclareMathAlphabet{\pazocal}{OMS}{zplm}{m}{n}
\newcommand{\calI}{\pazocal{I}}

\newcommand{\vz}{$v_0$~}

\newcommand{\nbrex}{neighbor exchange}
\newcommand{\nbrexs}{neighbor exchanges}
\newcommand{\ti}{T1 process}
\newcommand{\tis}{T1 processes}
\newcommand{\tti}{$\tau_{T1}$~}
\newcommand{\dti}{$\tau_{w}$~}

\newcommand{\instdti}{$\bar{\tau}_{w}^{inst}(t)$~}
\newcommand{\mdti}{$\langle \tau_{w} \rangle$~}
\newcommand{\tratio}{$\tau_{w}/\tau_{T1}$~}
\newcommand{\mtratio}{$\langle \tau_{w} \rangle/\tau_{T1}$~}
\newcommand{\fs}{$F_s(\tilde{q},t)$~}
\newcommand{\ta}{$\tau_{\alpha}$~}
\newcommand{\tb}{$\tau_{\beta}$~}

\begin{document}
\title{Controlled neighbor exchanges drive glassy behavior, intermittency and cell streaming in epithelial tissues}   
\author{Amit Das}
\affiliation{Department of Physics, Northeastern University, MA 02115, USA}
\author{Srikanth Sastry}
\affiliation{Jawaharlal Nehru Centre for Advanced Scientific Research, Jakkur Campus, Bangalore 560064, India}
\author{Dapeng Bi}
\affiliation{Department of Physics, Northeastern University, MA 02115, USA}

\begin{abstract}
Cell neighbor exchanges are integral to tissue rearrangements in biology, including development and repair. Often these processes occur via topological T1 transitions analogous to those observed in foams, grains and colloids. However, in contrast to in non-living materials the T1 transitions in biological tissues are rate-limited and cannot occur instantaneously due to the finite time required to remodel complex structures at cell-cell junctions.  
Here we study how this rate-limiting process affects the mechanics and collective behavior of cells in a tissue by introducing this important biological constraint in a theoretical vertex-based model as an intrinsic single-cell property.
We report in the absence of this time constraint, the tissue undergoes a motility-driven glass transition characterized by a sharp increase in the intermittency of cell-cell rearrangements.  Remarkably, this glass transition disappears as T1 transitions are temporally limited. 
As a unique consequence of limited rearrangements, we also find that the tissue develops spatially correlated streams of fast and slow cells, in which the fast cells organize into stream-like patterns with leader-follower interactions, and maintain optimally stable cell-cell contacts.  
The predictions of this work is compared with existing in-vivo experiments in Drosophila pupal development.   
\end{abstract}
\maketitle

Cell neighbor exchange is fundamental to a host of active biological processes, from embryonic development~\cite{Irvine827,Walck-Shannon2013} to tissue repair~\cite{Carvalho4267,tetley2019tissue}. Often referred to as cell intercalations~\cite{Tada3897}, it is the leading mode of rearrangement in a confluent cell-packing, and in the simplest form described by a topological \ti~ observed in foams~\cite{weaire2001physics}. During development, T1 rearrangements lead to diverse reorganization patterns~\cite{lienkamp2012vertebrate,Shindo649,10.7554/eLife.07090,NISHIMURA20121084,rozbicki2015myosin}, based on polarized global cues~\cite{Bertet2004,Kasza11732}, or apparently random localized active fluctuations~\cite{HELLER2014617,CURRAN2017480}. Disruption of \nbrexs~ lead to defects in developing embryos~\cite{Bertet2004,Kasza11732}, while  an increase has been shown to alleviate disease conditions~\cite{lienkamp2012vertebrate,Nishio295}. It is, therefore, important that the underlying biological programs of neighbor exchanges are regulated in a tissue. 

The steps associated with a T1 event, namely the shrinkage and restoration of a cell-cell junction, are driven actively by myosin II motors, present inside the cell and at the junctions ~\cite{CURRAN2017480,Simoes575,10.7554/eLife.34586,Munjal2015,Kasza11732,Rauzi2010,FERNANDEZGONZALEZ2009736}. Different global and local signals drive this activity~\cite{Kasza11732,CURRAN2017480,10.7554/eLife.12094}, and perturbing them  can disrupt or enhance \nbrexs~\cite{Bertet2004,Kasza11732,CURRAN2017480}. 
However, not much is known about how these signals determine the rate of \nbrexs~ at the level of a single cell and how these effects influence multicellular behavior.

While there is a scarcity of quantitative measurement of cell \nbrex~ rates~\cite{CURRAN2017480}, recent evidences suggest that limiting the rate of \nbrexs~ influences many tissue level properties. The fluidity of an epithelium~\cite{IYER2019578} 
is affected by differential rates of \nbrexs~ in development~\cite{CURRAN2017480,doi:10.1098/rstb.2017.0328}.
Neighbor exchanges and dynamic remodeling of cell-cell contacts~\cite{Scarpa143} are
extremely crucial in numerous examples of collective cell migration through dense environments, examples include the migration of border cells in developing Drosophila oocyte~\cite{cai2014mechanical}, the formation of dorsal branches in Drosophila tracheae~\cite{shaye2008modulation}, the formation of Zebrafish posterior lateral line primordium~\cite{Revenu1282} and the remodeling of lung airway epithelium under asthmatic conditions~\cite{Park_NMAT_2015}. 
Invading cancer cells often form  multicellular streams to migrate through narrow spaces in rigid tissues~\cite{patsialou2013intravital,friedl2012classifying} 
that depends crucially on \nbrexs, as suggested recently~\cite{10.1093/intbio/zyz015}.

Despite these evidences, it is not clear how the \nbrexs ~and dynamic remodeling of cell-cell contacts are related, and what roles they play in an emergent multicellular behavior. In this work we try to understand this relation within a theoretical approach. We extend the well-known 2D vertex model~\cite{Nagai_PMB_2001,Farhadifar_CB_2007} which has already provided useful insights on morphogenesis~\cite{Fletcher_BJ_2014,Spahn_2013}, epithelial maturation~\cite{Park_NMAT_2015}, unjamming~\cite{Mitchel_unpublished}, and wound-healing~\cite{staddon_plos_2018}.
In all these cases, the T1 events have been considered instantaneous or implicit. Furthermore, the interactions of active cell motility and the rate of T1 events have not been considered before to the best of our knowledge. 

The focus of this work is to study the effect of cell motility and rate of T1 events regulated at the level of individual cells. We find that changing the intrinsic cell-level persistence time for T1 events induces a gradual slowing of the cellular dynamics. Though this  is reminiscent of the dynamical arrest in glassy systems, the nature of the glassy state that emerges is distinct and not previously described. The interplay of cell motility and intrinsic persistence of T1 events gives rise to out-of-equilibrium states that behave largely like a glassy system, yet lacks conventional dynamic heterogeneity and surprisingly contains a population of mobile cells that can migrate via an unusual coordinated stream-like motion. 

\section*{Results}
\begin{figure*}[htbp]
\begin{center}
\includegraphics[width=1.8\columnwidth]{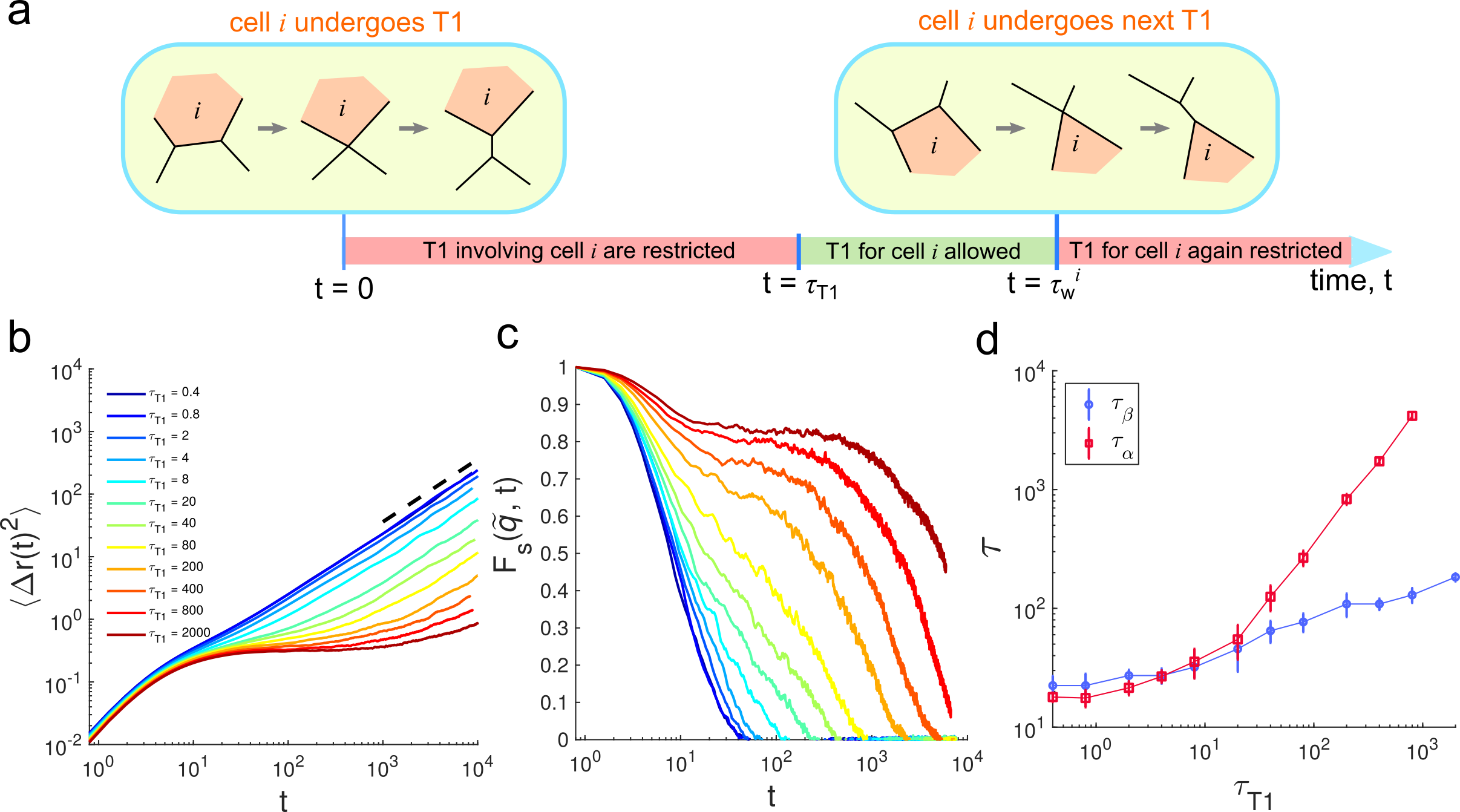}
\caption{
\textbf{Increased persistence of T1 events slows down cell dynamics.} (a) Schematic showing how we introduce the time-delay \tti between two successive T1 events associated with a given cell $i$. The waiting time $\tau_{w}^{i}$ between the two T1 events is always greater or equal to \tti due to the stochastic nature of cell-cell junction remodeling. (b) Mean square displacements (MSD) of the cell centers for fixed \vz$ = 1.6$ at different \tti. The dotted line has a slope of 1 
 on a log-log scale. (c) Self-intermediate scattering function \fs for the same \vz at different \tti. Here $\tilde{q}=\pi/\sqrt{A_0}$ where $\sqrt{A_0}$ is the unit of length in our model. (d) $ \beta $ and $\alpha$-relaxation timescales, \tb and \ta as functions of $\tau_{T1}$, respectively. 
}
\label{fig1}
\end{center}
\end{figure*}

\textbf{Model --} Our appropriately modified ``dynamic vertex model" (DVM)~\cite{Mitchel_unpublished} retains most of the classic features~\cite{bi_nphys_2015,Bi_PRX_2016}. Cells in a 2D epithelial monolayer are described by irregular polygons, defined using vertices, which constitute the degrees of freedom for the model. The vertex positions $\{ \bm{r}_i \}$ evolve against a uniform frictional drag $\zeta$ according to the over-damped athermal equations of motion
\be
\zeta \frac{d\textbf{r}_{i}}{dt} = \textbf{f}_{i}^{\text{shape}} + \textbf{f}_{i}^\text{active}. 
\label{eom}
\ee

Due to the biomechanical interactions cells resist changes to their shapes, described by the tissue mechanical energy~\cite{bi_nphys_2015,Farhadifar_CB_2007},
\be
E = \sum_{i=1}^N \left[ K_A (A_i-A_0)^2+ K_P (P_i-P_0)^2 \right]. \label{vm_eq}
\ee
where $N$ is the number of cells, $A_i$ and $P_i$ are the area and the perimeter of cell $i$, respectively. $A_0$ and $P_0$ are the equilibrium cell area and perimeter, respectively, considered uniform across the tissue~\cite{PhysRevLett.123.058101}. 
$K_A$ and $K_P$ are the elastic moduli associated with deformations of the area and perimeter, respectively. 
The second term in the right hand side of Eq.~\ref{vm_eq} yields a dimensionless target cell shape index $p_0 = P_0/\sqrt{A_0}$
~\cite{Park_NMAT_2015,Mitchel_unpublished}. 
Here we choose a constant $p_0 = 3.6$, which describes a solid-like tissue when cell motility is low~\cite{Bi_PRX_2016}. The force on any vertex due to cell shape fluctuations is
\be
\textbf{f}_{i}^{shape} = -\frac{\partial E}{\partial \textbf{r}_{i}} \label{shape_force}
\ee 

Each cell in the DVM behaves like a self-propelled particle with motility force \vz~\cite{Bi_PRX_2016,Yang_PNAS_2017} acting on the geometric cell center. The total active force on vertex $i$ is
\be
\textbf{f}_{i}^{active} = v_0 \tilde{\textbf{n}}_{i}   \label{active_force}
\ee
where $\tilde{\textbf{n}}_{i}$ is the direction of average active force on vertex $i$ 
(Fig.~\ref{figS1} and Methods). 

The most important ingredient in our model is that we allow \tis~ with a time delay \tti between successive T1 events (Fig.~\ref{fig1}a). For each cell, a timer $\delta t_{c}$ is kept which records the time elapsed since the last T1 transition involving the cell. Then all edges adjacent to the cell can undergo a T1 if and only if the length of the edge is shorter than a threshold $l_{min} = 0.1$ and $\delta t_{c} > \tau_{T1}$. Initially, cells are seeded with a random $\delta t_{c}$ chosen from a uniform distribution $[0,\tau_{T1}]$. This rule  introduces a persistent memory governing intercalations of cells.

\begin{figure*}[htbp]
\begin{center}
\includegraphics[width=1.6\columnwidth]{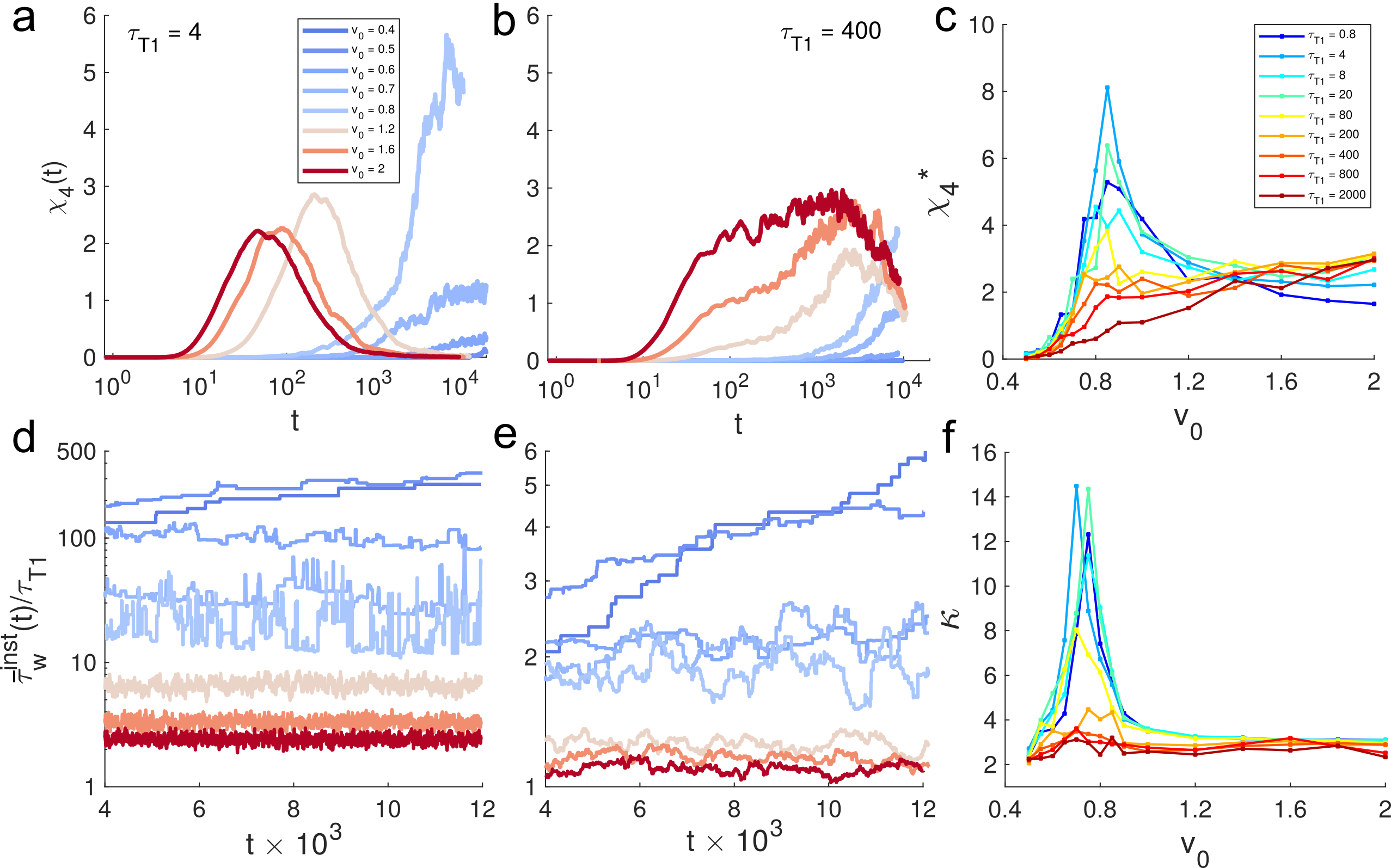}
\caption{
\label{fig2}
\textbf{Persistence in T1 events introduces variable degrees of dynamic heterogenity and drives intermittency of rearrangements.
} 
(a-b) Four-point susceptibility $\chi_{4}(t)$ for different \vz at two different \tti. 
(c) Peak value of $\chi_4$ denoted as $\chi_4^{\ast}$  vs. \vz.
(d-e) Time series of the instantaneous average of the waiting times of all cells, given by, $\bar{\tau}_{w}^{inst}(t)$, for the parameters corresponding to panels in a and b. 
(f) The kurtosis $\kappa$ of the observed time series of $\bar{\tau}_{w}^{inst}(t)$, as function of \vz. 
}
\end{center}
\end{figure*}

\textbf{ Dynamical slow down in cellular motion due to persistence in T1 events --}\label{cell_dynamics} 
According to previous results using similar models~\cite{Bi_PRX_2016,Yang_PNAS_2017}, solid-fluid transition in a motile tissue at $p_0 = 3.6$ happens for $v_0 > 0.6$. In order to understand the effects of single cell persistence time of T1 events, we start with a high motility fluid at this $p_0$. 
To quantify the cell dynamics we compute the mean square displacements (MSD), $\langle \Delta r(t)^2\rangle$ of the geometric centers of the cells (Fig.~\ref{fig1}b and~\ref{figS2}a) as well as the self-intermediate scattering function, $F_{s}(\tilde{q},t)$  (Fig.~\ref{fig1}c), a standard measure in glassy physics to quantify structural relaxation (see Methods for calculation details).
When \tti is small, the cell motion is nearly ballistic at short times and diffusive at long times ($\langle \Delta r(t)^2\rangle \propto t$) as expected in a typical viscoelastic fluid. This situation corresponds to nearly instantaneous T1 events typically considered previously~\cite{Farhadifar_CB_2007,Fletcher_BJ_2014,Park_NMAT_2015,staddon_plos_2018}.
Here the $F_s(\tilde{q},t)$ decays  sharply with a single timescale of relaxation indicating the tissue fluidizes quickly due to motility (Fig.~\ref{fig1}c).
Interestingly, as  \tti is increased the MSD starts to develop a `plateau' after the early ballistic regime and requires increasingly longer time to become diffusive. The development of this plateau is indicative of the onset of kinetic arrest and shows up as a two-step relaxation in terms of $F_s(\tilde{q},t)$ (Fig.~\ref{fig1}c). 
In glassy physics~\cite{Ediger_angell_nagel,Angell1924,Blaaderen1177,Weeks627}, the origin of the two-step relaxation has been attributed to  $\beta$-relaxation, taking place at intermediate times when each cell just jiggles inside the cage formed by its neighbors, which is also responsible for the plateau in MSD; and $\alpha$-relaxation, taking place at later times when a cell is uncaged and undergoes large scale motion to make the MSD rise after the plateau. We  extract \tb, the timescale of $\beta$-relaxation~\cite{PhysRevLett.116.085701} by locating the point of inflection in MSD curve on a log-log plot (see Methods and Fig.~\ref{figS2}b). The time scale \ta associated with $\alpha$-relaxation can be extracted from $F_s(\tilde{q},t)$ according to convention (Methods). In Fig.~\ref{fig1}d we show the dependence of \tb and \ta on $\tau_{T1}$. For low \tti    (fast T1's), the two timescales coincide indicating \tti has no discernible effect on the dynamics but starting at $\tau_{T1} \approx 20$ and onward the difference between the two timescales grow drastically. 
This gives a characteristic timescale where the hindrance of T1's causes effective caged motion and kinetic arrest in cells.  
As these behaviors are  reminiscent of the onset of glassy behavior in super-cooled liquids, the natural question is whether introducing a T1-delay merely provides another route leading to a conventional glassy state, similar to e.g. lowering the temperature? To answer this question, we quantify the dynamical heterogeneity in states suffering kinetic arrest due to T1-delay and compare with more conventional glassy states obtained by lowering $v_0$ at short $\tau_{T1}$. 

We measure the  four-point susceptibility $\chi_4(t)$ which is a conventional measure of dynamical heterogeneity in glassy systems~\cite{berthier_biroli_rmp,PhysRevLett.116.085701}. For nearly instantaneous T1's (Fig.~\ref{fig2}a), the tissue behaves like a conventional super-cooled glass as $v_0$ is decreased: $\chi_4(t)$ exhibits a peak which shifts towards larger times with decreasing $v_0$. Together with the increase in the peak magnitude of $\chi_4(t)$, these results indicate the lengthscale and timescale associated with dynamic heterogeneity become increasingly larger due to lowering $v_0$, which plays the role of an effective temperature~\cite{Bi_PRX_2016}. Using the peak value $\chi_4^{*}$ (Fig.~\ref{fig2}c), the glass transition can be located at $v_0\approx 0.8$. 
In contrast, the behavior of $\chi_4(t)$ for large large T1 delay is very different. Here, $\chi_4(t)$ develops a plateau over decades in time (Fig.~\ref{fig2}b), which is significantly broader than in the low \tti regime. Further, hindering T1's here has also reduced the dependence of dynamic heterogeneity on \vz and the peak value of $\chi_4^{*}$ (Fig.~\ref{fig2}c) no longer exhibits any peaks as function of $v_0$. This suggests that \vz has been replaced as the rate-limiting factor determining cell rearrangements, which are instead dominated by $\tau_{T1}$. This analysis thus reveals that while the high \tti regime becomes kinetically arrested similar to a super-cooled liquid, it does so in a manner distinct from 
lowering the temperature. Here, the effects of effective temperature ($v_0$) on dynamics get `washed away' and the characteristic timescale is mostly set by $\tau_{T1}$. Therefore, we essentially have a glassy state (with a $\tau_\alpha$ that can be directly controlled using $\tau_{T1}$) with a \emph{low degree of dynamic heterogeneity}.

\textbf{Intermittency in T1 events points to a dynamic regime distinct from a conventional glass --\label{T1_dynamics}} 
Since the origin of a growing dynamic heterogeneity in glasses is attributed to highly intermittent motion of individuals~\cite{Weeks627}, we explicitly investigate how intermittency of T1 rearrangements depends on \vz and $\tau_{T1}$. 
By tracking all neighbor exchanges we maintain a time dependent list $\{\tau_{w}^i\}$ illustrated in Fig.~\ref{fig1}a. We define an instantaneous waiting time \instdti (Fig.~\ref{fig2}d-e) by averaging the entries in this list at a given time $t$. 
We also consider distributions of the waiting times, $P(\tau_{w})$ (Fig.~\ref{figS3}).

When T1 rearrangements are nearly unconstrained and motility is high, \instdti exhibits steady fluctuations about a mean value that is slightly larger compared to the lower bound set by $\tau_{T1}$. As motility is lowered T1 occurrences become less frequent and \instdti increases overall. Near the glass transition, the dynamics becomes highly intermittent, and consequently, T1 events are separated by a broad distribution of waiting times (Fig.~\ref{figS3}) and multiple T1 events can take place simultaneously akin to avalanches~\cite{berthier_biroli_rmp}. Deeper in the glass phase ($v_0 < 0.8$) we see \instdti 
increases slowly over time, reminiscent of the aging behavior observed in glassy materials ~\cite{Weeks627,berthier_biroli_rmp}. The intermittent fluctuations are dampened significantly as T1 delay becomes very large (Fig.~\ref{fig2}e).

\begin{figure*}[htbp]
\begin{center}
\includegraphics[width=1.6\columnwidth]{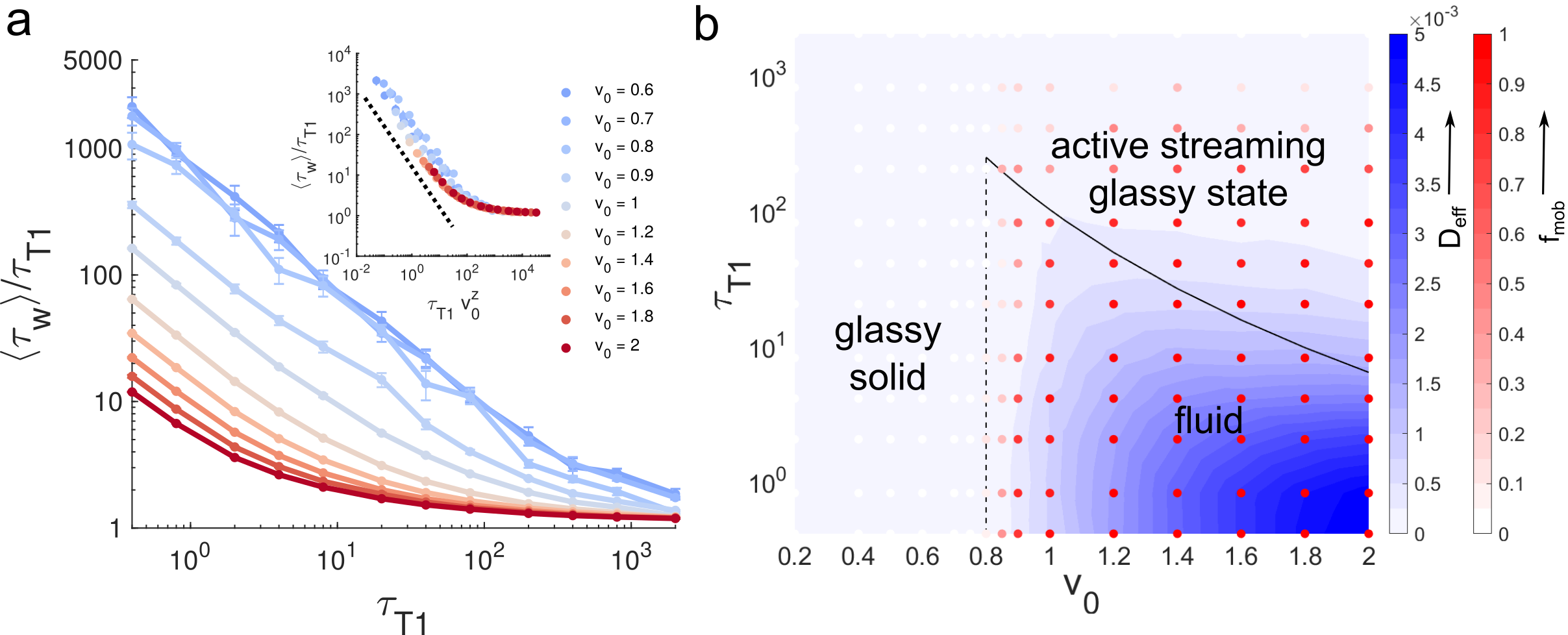}
\caption{
\textbf{Universal scaling of waiting times and phase diagram of cell dynamics.}
(a) Mean waiting times, scaled by $\tau_{T1}$, $(\tau_{w}/\tau_{T1})$ as function of \tti at different \vz. Inset, scaling collapse of \tratio with the scaling variable $x = \tau_{T1}v_{0}^z$ and exponent $z = 4.0 \pm 0.1$. The black dotted line is a guideline for power law of $x^{-1}$. 
(b) Phase diagram on \vz-\tti plane shows three different regimes of cell dynamics along with approximate phase boundaries. The heat map represents the values of effective diffusivity $D_{eff}$, and the intensities of the color of overlaid circular symbols represent the values of $f_{mob}$, the fraction of mobile cells with net displacements of at least 2 cell diameters. 
}
\label{fig3}
\end{center}
\end{figure*}

To further characterize the nature of fluctuations  and the intermittency, we compute the kurtosis $\kappa$ of the \dti distributions. 
In the small \tti regime, the distribution $P(\tau_{w})$ decays with power-law tails for \vz$<1$ (Fig.~\ref{figS3}), and this is also the location where $\kappa$ vs $v_0$ exhibits a very pronounced peak (Fig.~\ref{fig2}f) at the point of glass transition $(v_0 \approx 0.8)$. Away from the glass transition ($v_0>0.8$), $\kappa \approx 3$ corresponding to Gaussian fluctuations (Fig.~\ref{fig2}d,e and Fig.~\ref{figS3}). However, when $\tau_{T1}$ is increased, the intermittency fades away and the peak in $\kappa$ disappears.

\textbf{Universal scaling separates fast and slow regimes --\label{scaling}} Next we analyze the interplay between the T1 delay %
and the observed mean waiting times \mdti (Fig.~\ref{fig3}a). 
When \tti is small, the ratio \mtratio remains always larger than unity and varies by orders of magnitude depending on $v_0$. As the delay increases, %
\mtratio approaches unity and depends weakly on $v_0$. %
This behavior suggests the following universal scaling ansatz

\be
\frac{\langle\tau_{w}\rangle}{\tau_{T1}} = f(\tau_{T1}~v_{0}^{z}).
\label{scaling_ansats}
\ee  
In Eq.~\eqref{scaling_ansats}, $f(x)$ is the dynamical crossover scaling function with $x=\tau_{T1} v_{0}^{z}$. %

Re-plotting the data using Eq.~\eqref{scaling_ansats} we find good scaling collapse with exponent $z = 4 \pm 0.1$ (Fig.~\ref{fig3}a-inset). %
This uncovers two distinct regimes. %
For small $x$, $f(x) \sim x^{-1}$, which implies  \mdti$\propto v_0^{-z}$. %
We refer to this as the \textbf{fast} rearrangement regime. 
For large $x$, $f(x) \to 1$ %
indicating a \textbf{slow} rearrangement regime where \mdti$\propto \tau_{T1}$ and independent of $v_0$. 
The transition between the two regimes occurs at $x = x^*\approx 100$, corresponding to a scaling relation of $\tau_{T1} = x^* v_0^{-z}$ that constitutes the boundary separating these two rearrangement regimes. 

\begin{figure*}[htbp]
\begin{center}
\includegraphics[width=2\columnwidth]{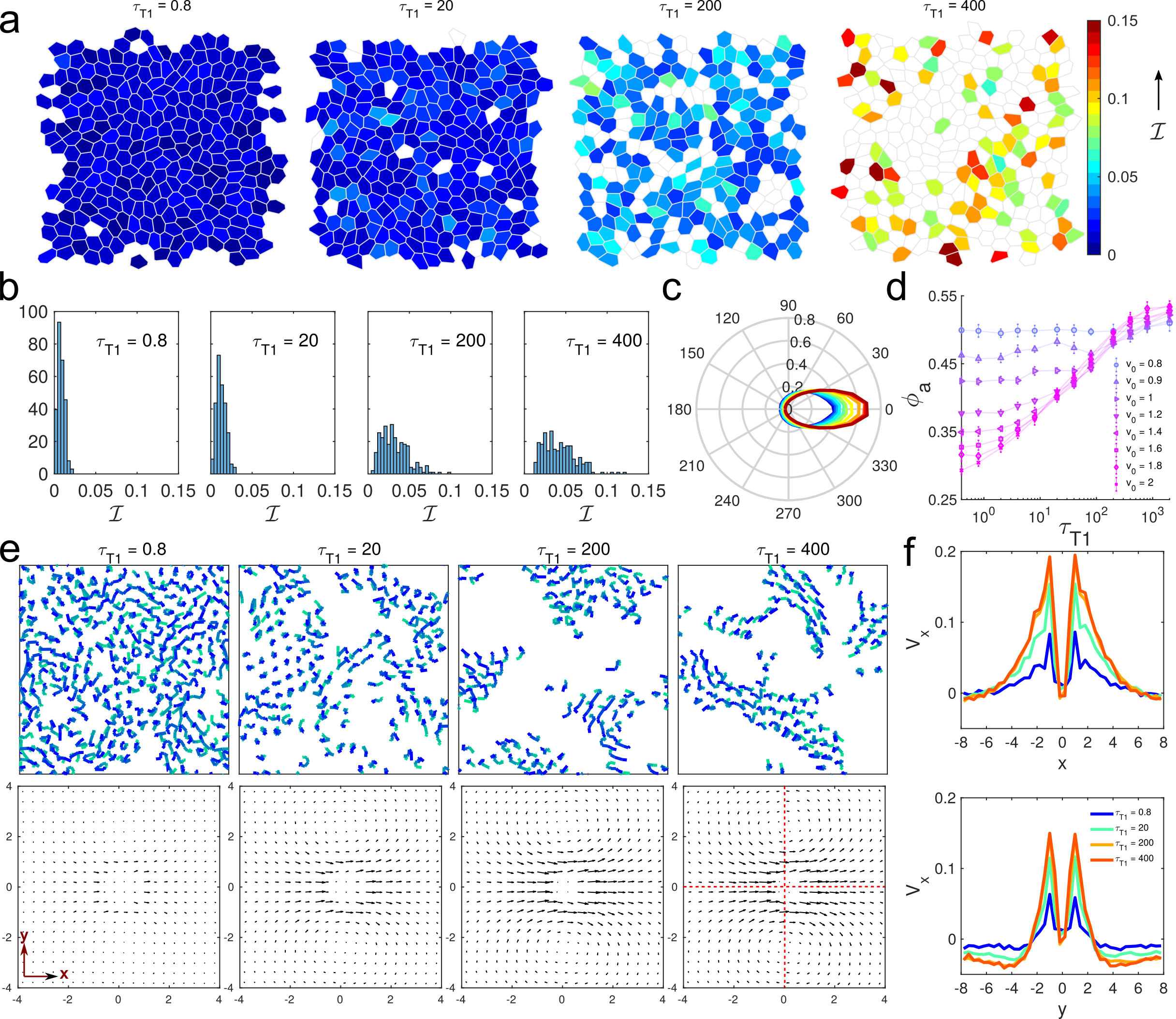}
\caption{
\textbf{Fast cells organize into streams as T1 time-delay increases.} 
(a) Simulation snapshots for four different \tti values at fixed \vz$ = 1.6$. Only the fast cells are shown (net displacement $d_{\infty} \gtrsim 2$ cell diameters, also see methods) are color-coded according to their intercalation efficiencies, defined as, $\calI = d_{\infty}/n_{T1}$ where $n_{T1}$ is the net T1 count. (b) Distributions of cell-level $\calI$ values, for different $\tau_{T1}$. (c) Distributions of angles (shown as polar plots) between displacement vectors of a pair of cells separated by less than $2$ cell diameters. This is calculated for cells with $\calI \geq \bar{\calI}$, the mean intercalation efficiency per cell in the tissue. 
(d) Average cell alignment probability $\phi_{a}$ as function of \tti for different \vz. 
The error bars are standard deviations of the mean for 10-20 samples.
(e) The individual cell trajectories at different \tti and fixed \vz$ = 1.6$. Top row - cell trajectories for a short time period. Colors change from green to blue to mark cell positions progressively forward in time. Trajectories are shown only for the cells that travel 80\% of a cell diameter or more distance in this time window. Bottom row - average velocity field $\Vec{V}$ around any cell corresponding to each panel in top row. 
Here the reference cell is always at the origin, its velocity vector pointing from left to right. (f) Plots of $V_x$, the x-component of $\Vec{V}$, 
 along two lines parallel to x- (top) and y-axis (bottom), shown as dotted lines in the rightmost panel of the bottom row in e.
}
\label{fig4}
\end{center}
\end{figure*}

The dynamics in the \textbf{fast} regime is not hindered by $\tau_{T1}$ and is largely driven by the effective temperature $T_{eff} \propto v_0^2/D_r$~\cite{Bi_PRX_2016}. For high values of $v_0$ in this regime, the motile forces are sufficient to overcome the energy barrier associated with T1 transitions; however as $v_0$ drops below $v_0^* \approx 0.8$, the tissue enters into a \textit{glassy solid} state where cell motion become caged. 

In the \textbf{slow}  regime, $\tau_{T1}$ dominates the dynamics as it is the longest timescale in the system and therefore the ultimate bottleneck to rearrangements. This leads to \mdti to depend linearly on \tti while being insensitive to motility. 
In addition to glassy-behavior which is a source of non-equilibrium fluctuations, here \tti constitutes another possible route that can take the system out-of-equilibrium, effectively slowing down the dynamics. 

Here, the uncaging timescales ($\tau_\alpha$) grows with \tti (Fig.~\ref{fig1}d) but remain quite finite, as in a highly viscous or \textit{glassy fluid}. %

\noindent
\textbf{Phase diagram of cellular dynamics --\label{phase_diagram}} 
A summary of these results suggest that the phase diagram on the \vz$-$ \tti plane (Fig.~\ref{fig3}b) can be categorized into three phases: 
 \textbullet~ glassy solid, \textbullet~ fluid, and an unusual \textbullet~ \emph{active streaming glassy state (ASGS)}, to be discussed in depth below. 
The solid-fluid phase boundary is given by peak-positions in $\kappa$ (Fig.~\ref{fig2}f) and $\chi_4^{\ast}$ (Fig.~\ref{fig2}c), and effective diffusivity, $D_{eff}$ (Methods, Fig.~\ref{fig3}b, and Fig.~\ref{figS2})~\cite{Bi_PRX_2016}. 
The boundary between fluid and the ASGS phase is given by the crossover scaling (Eq.~\ref{scaling_ansats}) between the \textbf{slow} and \textbf{fast} regimes. 

The ultra-low $D_{eff}$ values in the ASGS phase resemble a solid more than a fluid. However the finite \ta (Fig.~\ref{fig1}d) and the nature of waiting times (Fig.~\ref{figS2}) indicate that these states are only solid-like for timescales less than \tti. 
Furthermore, it appears that the eventual transition to diffusive motion (at times $ > \tau_{T1}$) occurs in a highly heterogeneous manner and the motion is dominated by a small population of \emph{fast moving cells}. To understand this peculiar nature of fluidity we quantify the fraction of relatively fast moving or mobile cells, given by $f_{mob}$ (Methods). 
 As expected, $f_{mob}$ is large and unity for \textit{fluid} and zero for \textit{glassy solid}. However, in the ASGS phase, although $f_{mob}$ decreases slowly as \tti increases,  curiously enough, it remains finite. These evidences suggest that the ASGS phase always includes states with a {\bf distribution of fast and slow cells}. This could be signature of another kind of growing heterogeneity %
which we try to understand next. 

\begin{figure*}[htbp]
\begin{center}
\includegraphics[width=1.8\columnwidth]{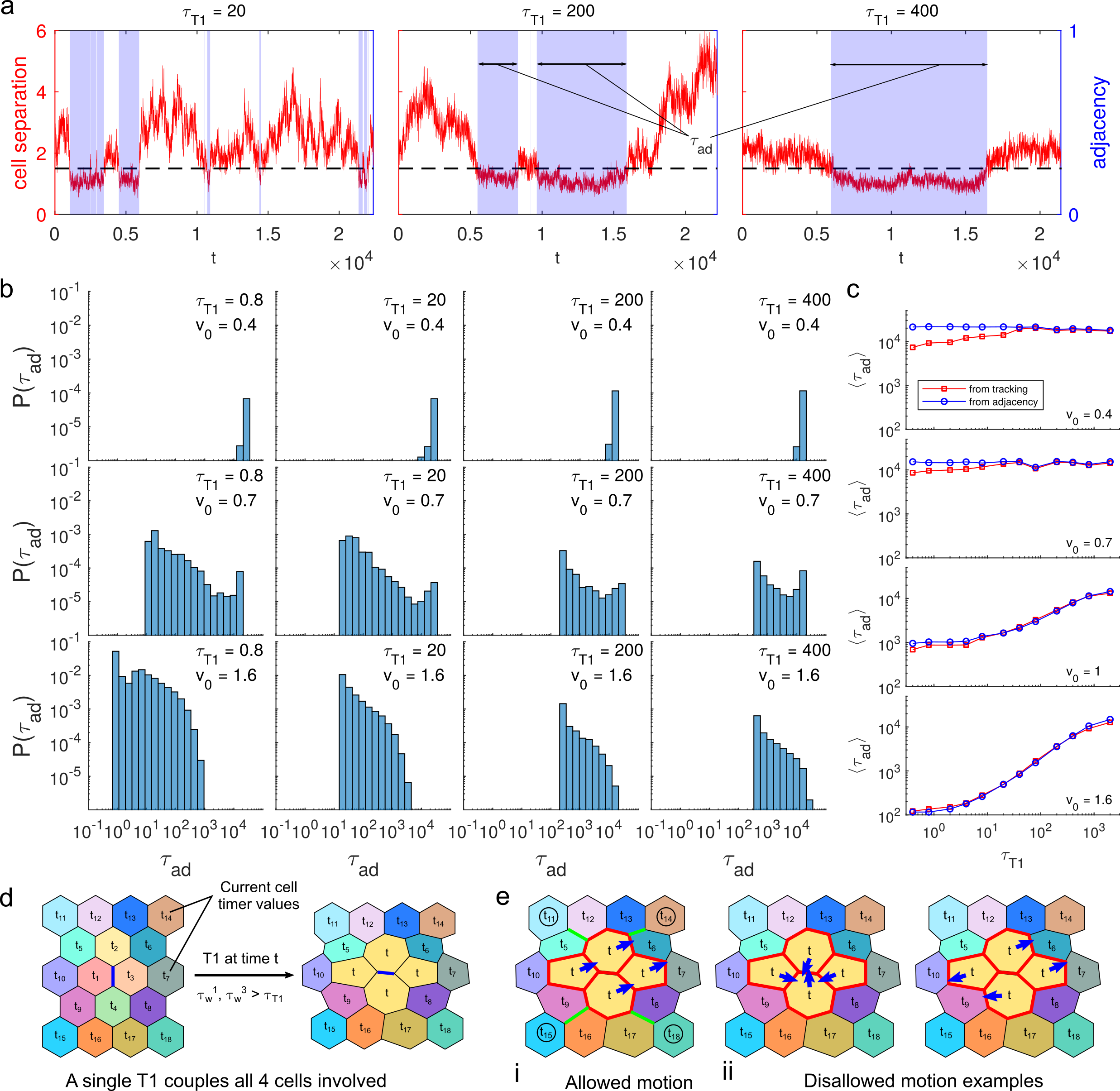}
\caption{
\textbf{Effective cell-cell cohesion depends on T1-delays and introduces cell streaming.} (a) Typical time evolution of the distance separating two cells. Blue shaded areas indicate durations for which they remain touching, defined as when their distance is below a threshold of 1.5. Examples are shown for different \tti at fixed \vz=1.6.
A single pair of cells can undergo multiple events of cohesion. We refer to the time duration of these events as effective adhesion timescales $\tau_{ad}$ 
(b) Distributions of $\tau_{ad}$, at different \vz and \tti. (c) Mean $\tau_{ad}$, estimated from cell adjacency information (blue squares) and tracking of cell centers (red squares), as a function of \tti at different \vz. (d-e) Proposed mechanism of how effective cohesion can lead to cell streaming. Panel d shows a tissue patch of 18 cells each with a different timer $\delta t_{i} = t_i$ recording the time elapsed since last T1. Cells 1 and 3, (sharing the blue colored junction) undergo T1 at time $t$ as their respective waiting times exceed $\tau_{T1}$. This T1 not only resets the timers on cells 1 and 3, it also resets the timers on nearby cells 2 and 4 because they now become neighbors after the T1. This resetting introduces an effective adhesion in several cell pairs and forbids another T1 event involving the junctions colored in red for a period at least $\tau_{T1}$. Panel e shows the consequence of this effect. These 4 cells can now only move coherently (case i) with types of motions forbidden, as illustrated here (case ii). Thus, this cluster of 4 cells act as a nucleator for a cell stream, that can grow in length as any of neighboring cells, marked by circles, can join the cluster via T1 transitions involving the junctions colored green (case i) and propagate the streaming. 
}
\label{fig5}
\end{center}
\end{figure*}

\noindent
\textbf{Efficient fast cells organize into cellular streams --\label{cell_streaming}} We observe that depending on the phase they belong to the fast cells exhibit different degrees of neighbor exchanges. 
To quantify this, we define a single cell intercalation efficiency $\calI$ (Fig.~\ref{fig4}a and Methods). For fast cells $\calI$ increase with increasing \tti at a fixed motility (Fig.~\ref{fig4}a). At the same time, the spatial distribution of the fast cells become increasingly heterogeneous, and stream-like. 
The distributions of $\calI$ (Fig.~\ref{fig4}b) also reflect this change, as they become wider and heavy-tailed, with 2-5 fold increase in the mean efficiency $\bar{\calI}$ as \tti goes up. The fast cells also exhibit mutual spatiotemporal alignment (Methods), which 
grows rapidly with \tti (Fig.~\ref{fig4}c). The overall alignment probability $\phi_{a}$ (Methods, Fig.~\ref{fig4}d) can be used as an order parameter for streaming behavior. At small motilities, $\phi_{a}$ is insensitive to \tti and large due to collective vibrations in a solid-like tissue~\cite{Bi_PRX_2016}. At higher motilities, $\phi_{a}$ becomes increasingly dependent on \tti. 
Taken together, these analyses pinpoint the necessary conditions for stream-like behavior: increasing T1 delay alone can induce alignments, but higher motilities are also crucial.

To visualize this streaming behavior we follow cell trajectories (Fig.~\ref{fig4}e-top), which are uniformly distributed and randomly oriented for low $\tau_{T1}$, but gradually become sparse and grouped into stream-like collectives as \tti increases. These collectives are highly correlated and persistent. 
Interestingly, here we have used a time interval much smaller than the corresponding $\beta$-relaxation timescales to compute the displacement vectors of cell centers. Therefore, this type of collective behavior occurs even before the cells uncage. 

To quantify the spatial correlations arising during streaming we adapt a quantity $\vec{V}$~\cite{Szab__2010}, computed in the co-moving frame of a given cell, which represents the average velocities at different locations around it. Collective migratory behavior shows up as vectors of similar length and direction near the cell, whereas solid or fluid-like behavior results in isotropic organization of vectors of uniform sizes. In Fig.~\ref{fig4}e-bottom, signatures of collective motions are absent when \tti is small, but gradually appears with increase in \tti as the sizes and orientations of the vectors become more correlated along the direction of motion and remain uncorrelated along the direction perpendicular to it. Such anisotropic vector fields are hallmarks of cellular streaming~\cite{Szab__2010} that involves high front-back correlations (Fig.~\ref{fig4}f-top), and low left and right correlations (Fig.~\ref{fig4}f-bottom). 
This also provides the size of a typical stream, about 8 cells long and 4 cells wide. We further find that these streams are driven by leader-follower interactions~\cite{das2015molecular,doi:10.1091/mbc.e16-05-0329} that increase with \tti (Fig.~\ref{figS4}). 
In this light, after quantifying the streaming behavior, we consider its possible underlying mechanisms. One big candidate is dynamic remodeling of cell-cell contacts.

\begin{figure}[htbp]
\begin{center}
\includegraphics[width=1\columnwidth]{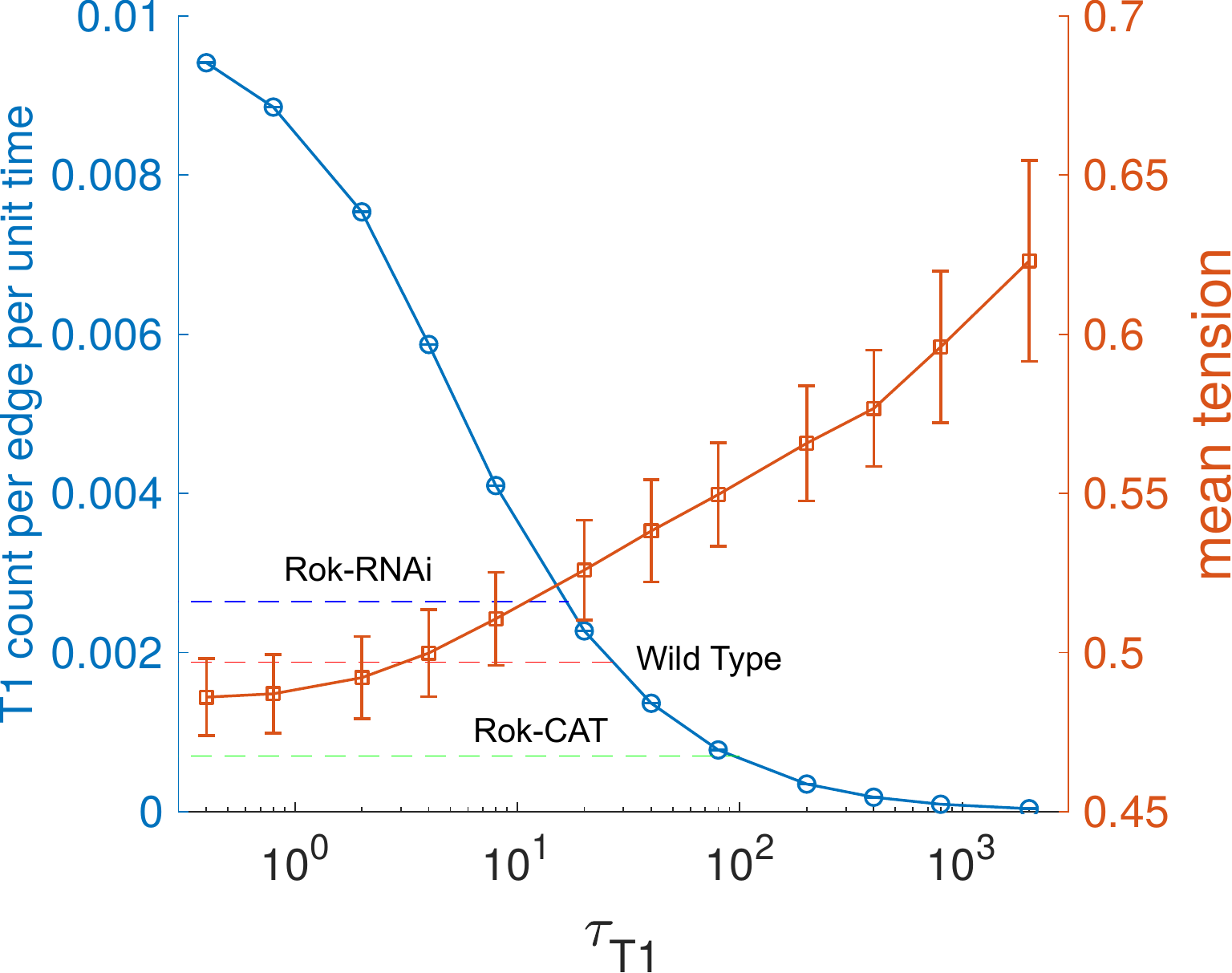}
\caption{
\textbf{Prediction of T1 rates and comparison with experimental data on Drosophila pupa development.} Mean rate of T1 events and mean tension, calculated per junction from our simulations at \vz$= 1.6$ as function of \tti.  The dashed lines represent the experimentally measured rates of neighbor exchange in Drosophila pupal notum under three situations from Ref.~\cite{CURRAN2017480}: the wild type notum (red), with over-expression of a constitutively active Rho-kinase (Rok-CAT, green) that enhances junctional myosin-II activity and hence increases tension, and with a Rho-kinase RNAi (Rok-RNAi, blue) that reduces junctional myosin-II activity and hence decreases tension. The observed changes in T1 rate due to these two perturbations are consistent with our predicted relationship between junctional tension and T1 rates.
}
\label{fig6}
\end{center}
\end{figure}

\noindent
\textbf{Delayed T1 events result in effective cell-cell cohesion --\label{adhesion}} 
The delays between successive T1 events in our model essentially provide a dynamical way to control stability of cell-cell contacts. Two neighboring cells can maintain an effective adhesion for a time-period equal to \tti or longer, depending on the time-evolution of the shared junction. 
In simulations, we have direct access to the characteristic time of this effective adhesion, $\tau_{ad}$, which can be obtained from the cell-cell adjacency information. Alternatively, a more experimentally accessible measurement can be performed using cell tracking. 
Fig.~\ref{fig5}a illustrates quantification of cell-cell cohesion in our simulations using both techniques.  
 When \tti is small, cell-cell contacts are frequent but short lived compared to the large \tti case for which the contacts become 
long-lived. The distributions of $\tau_{ad}$'s (Fig.~\ref{fig5}b) are very narrow for low motilities, with a single peak at $\tau_{ad} \approx \tau_{total}$, the total simulation time, as expected in the glassy solid state. Increase in motility widens the distributions and introduces a new peak at $\tau_{ad} \approx \langle\tau_{w}\rangle$.
At even higher motility, the peak at $\tau_{total}$ disappears. The new peak shifts to smaller values but occurs at $\tau_{ad} > \tau_{T1}$ in the fluid phase, and at $\tau_{ad} = \tau_{T1}$ in the ASGS. The mean adhesion times $\langle\tau_{ad}\rangle$ also reflect these trends (Fig.~\ref{fig5}c) and the two methods of extracting $\tau_{ad}$'s agree nicely. An important feature is that the variance of $\tau_{ad}$ is always large for high motility states, indicating a heterogeneity in the remodeling of cell-cell contacts. In the ASGS, $\langle\tau_{ad}\rangle$ has a regime of strong dependence on \tti (Fig.~\ref{fig5}c, bottom), 
indicating tunable window of cell-cell adhesions within the ASGS. 

The emergence of cell streaming from the effective cell-cell cohesion is an amplified tissue level response to the persistence memory introduced by the T1 delay. This can be shown by inspecting the consequences of a single T1 event (Fig.~\ref{fig5}d). It not only resets the timers on the two cells swapping the shared junction, but also resets timers on nearby cells that now become neighbors after the swap. These events introduce a cohesive 4-cell unit that is stable for a time period of at least $\tau_{T1}$. This time-restriction also sets a natural lower bound to the $\tau_{ad}$'s, as reflected in Fig.~\ref{fig5}b. A consequence of this effective cohesion is that these 4 cells can now only move coherently for a period $\tau_{T1}$ or more (Fig.~\ref{fig5}e). The only way this cluster can move is via T1 transitions involving junctions at the periphery, as illustrated in Fig.~\ref{fig5}e-i. This also allows this cluster to grow into a stream only when \tti is optimal, because very frequent or rare T1 events would not be useful. That is why we see a drop in $f_{mob}$ (Fig.~\ref{fig3}b) and saturation of $\tau_{ad}$ (Fig.~\ref{fig5}c) as \tti become very high. The movements of such streams are always unidirectional at any given timepoint which automatically introduces a leader-follower interaction among these cells, characterized in Fig.~\ref{figS4}. 

The above analysis also reveals that different glassy regimes (Fig.~\ref{fig3}b) can be distinguished based on measuring the characteristic time of cell-cell cohesion and its statistical distribution $P(\tau_{ad})$. For example, consider two different glassy states, one with a large $\tau_{T1}$ and one with small motility. In terms of conventional measures both states would exhibit glassy features (vanishing $D_{eff}$, large $\chi_4$), however, our work predicts that the state with a large $\tau_{T1}$ should have a broadly distributed $P(\tau_{ad})$ compared to the motility driven glass transition.

\noindent
\textbf{Predictions for Drosophila pupa development --}\label{Drosophila_predictions} Cell intercalations in our model are spatially uncorrelated with no directional polarity. Such unpolarized cell intercalations have been discovered recently in Drosophila notum in the early pupal stage~\cite{CURRAN2017480}. Evidences indicate that random fluctuations in junction lengths, strongly correlated with activity of junctional myosin-II, drive such unpolarized intercalations. Rate of intercalation drops as the pupa ages, and at the same time there is an increase in junctional tensions. We observe a similar decrease in the rate of T1 events in the model as \tti increases, accompanied by an increase in junctional tension (Fig.~\ref{fig6}) (Methods), consistent with the experimental observation. 
We can also compare the observed trends of experimental perturbations to the wild-type tissue with our results and predict that increasing or decreasing \tti in our model influences T1 rates in a manner similar to perturbing junctional tension by up-regulating or inhibiting junctional myosin-II activity, respectively.

\section*{Discussion and conclusion}\label{discussions}

To model collective dynamics of confluent epithelial cells, we introduced an inherent timescale ($\tau_{T1}$) for cells to undergo rearrangements based on the important observation that T1 events in real tissues do not occur instantaneously.   
When $\tau_{T1}$ is short compared to other timescales in the model, we recover a motility-driven glass transition, occurring at motilities large enough to overcome the energetic barriers that cause a cell to become caged. Near this glass transition, we observe highly intermittent cell motion, as well as \nbrexs~ events concomitant with growing dynamical heterogeneity. 

However, when $\tau_{T1}$ grows we discover a rich dynamical regime where the system appears glassy on timescales governed by $\tau_{T1}$ but become fluidized at longer times. Compared to the motility-driven glass transition, this regime has a completely different kind of heterogenous glassy behavior characterized by disappearance of conventional glass-fluid boundary and appearance of spatially distributed pockets of \textbf{fast} and \textbf{slow} cells. The origin of this new glassy behavior stems from the effective cell-cell cohesion caused by the inability to undergo local cellular rearrangements. Surprisingly, this local frustration actually serves to enhance collective migration by facilitating stream-like patterns, reminiscent of leader-follower behavior, albeit without any explicit alignment interactions between cells. Interestingly, this connection diminishes as mean effective adhesion times saturate when persistence of rearrangements becomes too low or too high (Figure~\ref{fig5}c). Therefore, the cell-cell contacts need to be dynamic, but optimally stable to maintain the streaming mode. Note that the stream-like motion we observe occurs in the absence of any dynamic heterogeneity and this sets it apart from string-like cooperative motion observed in a 3D supercooled liquid~\cite{Glotzer_string_supercooled} which is a direct consequence of dynamic heterogeneity.

This result, consistent with the existing biological mechanisms~\cite{friedl2012classifying,friedl2017tuning}, sheds new light on the nature of cell-cell adhesion associated with cell streaming in cancerous tissues~\cite{friedl2012classifying}. 

The effective adhesion picture that emerges from our study is also consistent with the recent observations regarding polarized intercalations in extending germ-band of Drosophila embryo~\cite{10.7554/eLife.34586}. Oscillations in adherens junction protein E-cadherin has been deemed necessary for successful intercalations, and inhibition of this oscillation led to a decrease in successful T1 events. Our predictions reverse engineer this effect by limiting the rate of T1 events which leads to an increase in effective cell-cell adhesion and stability of junctions. Increasing the E-cadherin levels at the junctions would be the biochemical way of achieving this as shown before~\cite{10.7554/eLife.34586}.

The unusual nature of the \textit{active streaming glassy state} has multitude of implications. The slow but finite structural relaxation gives the material a tunable viscosity which can be extremely useful for preparation of biology-inspired sheet-like objects of controllable stiffness~\cite{toda2018programming,nguyen2018engineered}. The control of T1 rate can be potentially translated to gene-level control of the signaling~\cite{xie2018designing} associated with developmental events and disease conditions that strongly depend on cell intercalations, \textit{e.g.} body-axis extension and kidney-cyst formation~\cite{Walck-Shannon2013}, respectively. This might even allow design of organisms with programmable development~\cite{teague2016synthetic}, or one with controlled disease-spreading rates.

\section*{Acknowledgments}
We acknowledge the support of the Northeastern University Discovery Cluster and the Indo-US Virtual Networked Joint Center project titled “Emergence and Re-modeling of force chains in soft and Biological Matter No. IUSSTF/JC-026/2016.

\section*{Methods}\label{methods}
\noindent
\textbf{Active cell motility in DVM --} The explicit form of the active force on any vertex $i$ depends on the motility force contributions from the adjacent cells:
\be
\tilde{\textbf{n}}_{i} = \sum_{c \leftrightarrow i} a_{c}\textbf{n}_{c} \label{vertex_polarization}
\ee
where $a_c = l_c/(2z_{i}\sum_{c \leftrightarrow i} l_c)$ is the weight associated with cell $c$ adjacent to the vertex $i$, $l_c$ is the total length of the edges shared by vertex $i$ and cell $c$, $z_{i}$ is the connectivity at vertex $i$, and the factor 2 takes care of double contributions from the same cell. This averaging scheme is different than the recent approaches using a flat average of active forces on adjacent cells~\cite{czajkowski2019glassy,Okuda2015,Park_NMAT_2015} and ensures that the active force on vertex $i$ is dominated by the cell sharing the longest edges attached to $i$. The polarization of the self-propulsion on  cell$-c$ is given by $\mathbf{n}_c = (cos~\theta_{c} \,, \ sin~\theta_c)$, where the polarization angle $\theta_c$ is perturbed only by a white-noise ~\cite{RevModPhys.88.045006,doi:10.1146/annurev-conmatphys-031214-014710,C3SM52469H,fily_marche_prl_2012,PhysRevE.77.046113}:
\be
\frac{d\theta_c}{dt} = \zeta_\theta
\ee 
where the $\zeta_\theta$ is a Gaussian white noise with zero mean and variance $2D_r$ which sets the repolarization timescale for the cells in our model given by $1/D_r$. We use a fixed $D_r$ for all cells throughout the present study. 

\noindent
\textbf{Simulation details --} Our simulations are overdamped dynamics of 256 cells periodic boundaries along both $x$ and $y$ directions. We use $\sqrt{A_0}$ as the unit of length, $K_P A_0$ as the unit of energy and $\zeta/K_P$ as the unit to measure time $t$ in our simulations. The vertex positions are updated by solving Eq.~\ref{eom} using Euler's scheme. Our dynamical simulations are initialized from random Voronoi configurations which have been subjected to energy minimization using the conjugate-gradient algorithm. All simulations have been done in the Surface Evolver program~\cite{Brakke} with a fixed equilibrium cell area $A_0 = \bar{A} = 1$ ($\bar{A}$, the mean cell area), time step of integration $\Delta t = 0.04$. We run each of our simulations for $\sim 10^6$ steps and collect data 
for subsequent analyses after the tissue properties like the mechanical energy reaches steady state. We scan the following parameter space: \vz $\in [0.2,2]$ and \tti $\in [0.4,2000]$ at a fixed $D_r = 0.5$. For each combination of \vz~ and \tti we perform 10-20 independent simulations. 

\noindent
\textbf{Measuring junctional tension --} Tensions arise in the vertex model due to mismatch between the actual cell perimeters and the equilibrium perimeter. It can be defined in terms of the preferred scaled perimeter or target cell shape index $p_0$. We calculate tension on any junction shared between cells $i$ and $j$ by 
\be
T_{ij} = (p_i - p_0) + (p_j - p_0)
\ee
where $p_i$ and $p_j$ are respective scaled cell perimeters, measured during the simulation. 

\noindent
\textbf{Determination of $\tau_{\beta}$ --} We extract the $\beta$-relaxation timescale from any given MSD vs $t$ plot by locating the minimum in the time derivative of MSD given by: $\frac{dln(\langle\Delta r^2(t)\rangle)}{dln(t)}$~\cite{PhysRevLett.101.267802}. Plots of this time derivative for MSDs shown in Fig.~\ref{fig1}b are shown in Fig.~\ref{figS2}b. 

\noindent
\textbf{Analysis of self-intermediate scattering function $F_{s}(\tilde{q},t)$ --} We have used the following definition of self-intermediate scattering function: $F_{s}(q,t) = \langle e^{\mathit{i}\mathbf{q} \cdot \Delta \mathbf{r}(t)}\rangle$ where $q$ is the wave vector corresponding to our length scale of choice and the angular brackets represent ensemble average and averages over angles made by $\mathbf{q}$ and $\Delta \mathbf{r}(t)$, cell displacement vectors for time-delay $t$. Conventionally, emphasis is given to the behavior at $q=2\pi/\sigma$, where $\sigma$ is the inter-particle particle separation for configurations with particles just touching. However, we choose to focus on a even more restrictive wave vector $\tilde{q}=\pi/\sigma$, corresponding to a length scale of 2 cell diameters. The reason for this is a technical one: since the degrees of freedom in the DVM are the vertices rather than the cell centers, the cell centers (calculated at every step based on vertices) can exhibit unusual fluctuations even when cells are completely caged. The current choice of $\tilde{q}$ allows us to consider relaxations where these artificial fluctuations contribute much less. To eliminate contributions from any local or global drift we consider temporal changes in nearest-neighbor separations as $\Delta \mathbf{r}(t)$, instead of pure displacements of cell centers.

We define the $\alpha$-relaxation timescale \ta as follows: $F_s(\tilde{q},\ t = \tau_{\alpha}) = 0.2$, following the definition used recently by another cell-based model study on similar tissues~\cite{refId0}.

\noindent
\textbf{Definition of quantities associated with mobility --} We have used several order parameters to describe different regimes of cell dynamics in our model tissue. Below we define them one by one. 
We define effective diffusivity by: $D_{eff} = D_s/D_0$ where $D_s = \lim_{t \rightarrow \infty}\langle \Delta r(t)^2 \rangle/4t$, the long time self-diffusion coefficient and $D_0 = v_0^2/2D_r$, the free diffusion coefficient of a single isolated cell. $D_s$ has been computed using the value of mean-square displacement at the maximum time delay allowed in our simulation.

To define the fraction of mobile cells $f_{mob}$, we follow individual cell MSD and define the net displacement of a cell: $d_{\infty} = \lim_{t \rightarrow \infty}\sqrt{\langle \Delta r^{2}(t) \rangle}$. Then we find out the number of cells $N_{mob}$ that have $d_{\infty}\gtrsim 2$ cell diameters. Finally, $f_{mob} = N_{mob}/N$. This definition is consistent with the definition of mobile particles used in Ref.~\cite{Kob_dynamic_het}.

We define individual cell intercalation efficiency as $\calI = d_{\infty}/n_{T1}$ where $n_{T1}$ is the net T1 count for the given cell in the entire simulation. 

\noindent
\textbf{Analysis of orientation alignment in cell trajectories --} To capture the orientational order and spatial organization of fast cells, we concentrate on the cells with intercalation efficiency $\calI \geq \bar{\calI}$, the mean intercalation efficiency. This gives us a list of fast cells. Then we consider the whole simulation trajectory and calculate the probabilities of the angle, $\theta_a$ between the instantaneous displacement vectors of any cell pair chosen from our list of fast cells, where the cell-cell separation is less or equal to 2 cell diameters. We consider all possible cell pairs satisfying this criterion and pool all the $\theta_a$ to generate the probability density function $P(\theta_a)$. 
We define the net alignment probability $\phi_{a}$, by the following:
\be
\phi_a = \int_{-t_c}^{t_c} P(\theta_a)d\theta_a
\ee
where $t_c = 30^{\circ}$.

\bibliography{dvm_glassy_fluid}

\clearpage
\setcounter{figure}{0}    
\renewcommand{\thefigure}{S\arabic{figure}}
\onecolumngrid
\section*{Supplementary Data}
\begin{figure*}[htbp]
\begin{center}
\includegraphics[width=0.2\columnwidth]{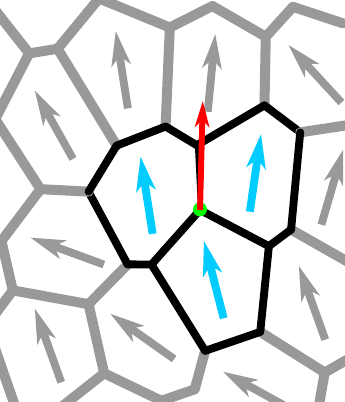}
\caption{
\textbf{Details of the DVM.} (a) Schematic of the dynamic vertex model showing how motion of any vertex depends on motion of its adjacent cells. 
}
\label{figS1}
\end{center}
\end{figure*}

\begin{figure*}[htbp]
\begin{center}
\includegraphics[width=0.6\columnwidth]{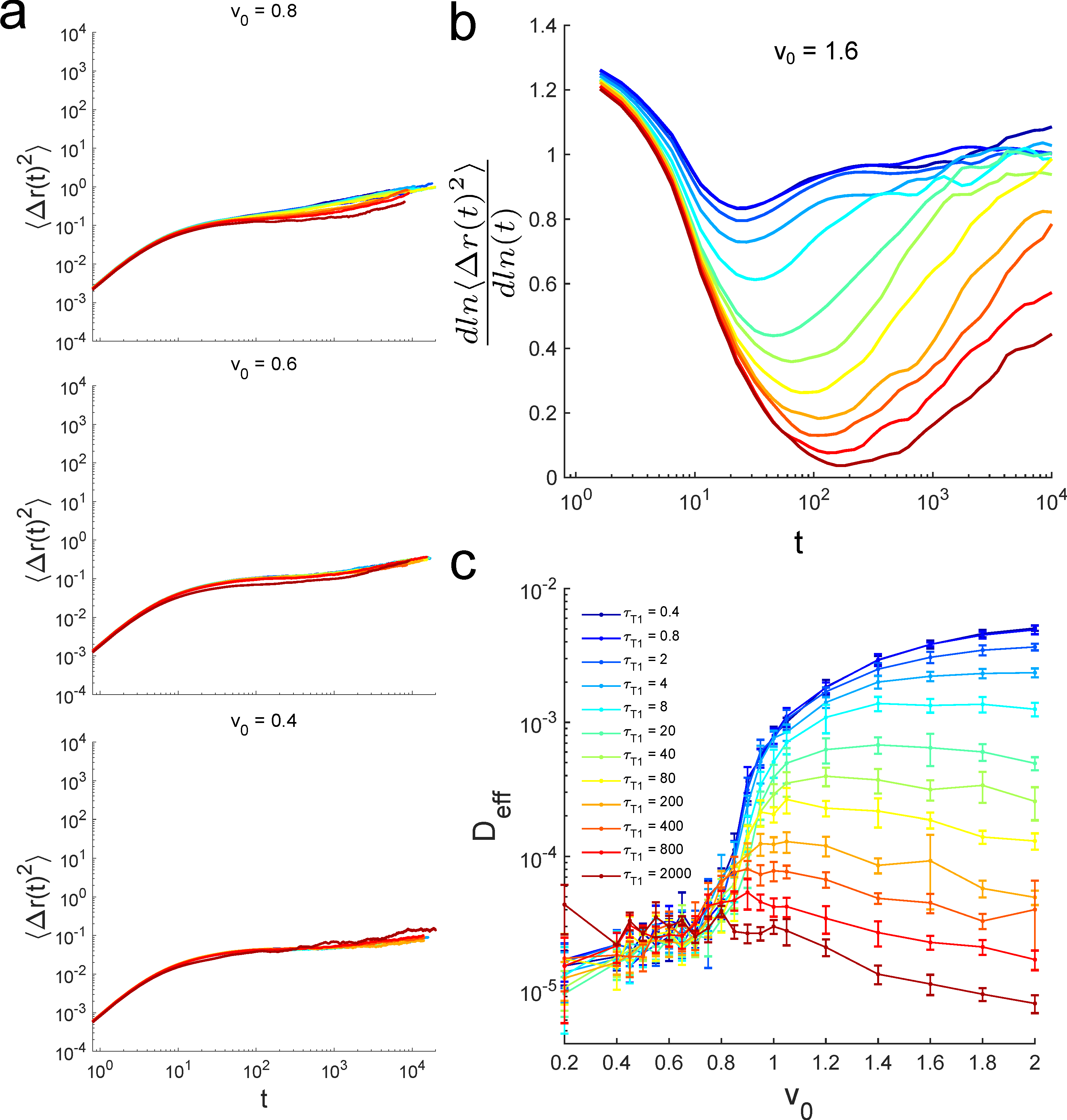}
\caption{
\textbf{Mean square displacements and associated analyzed dynamical quantities.} (a) Mean square displacements of the cell centers for three different \vz at different T1 time-delays \tti. (b) Time derivative of MSD for \vz=1.6. (c) Effective diffusivity $D_{eff}$ as function of \vz at different \tti.
}
\label{figS2}
\end{center}
\end{figure*}

\newpage
\begin{figure*}[htbp]
\begin{center}
\includegraphics[width=0.8\columnwidth]{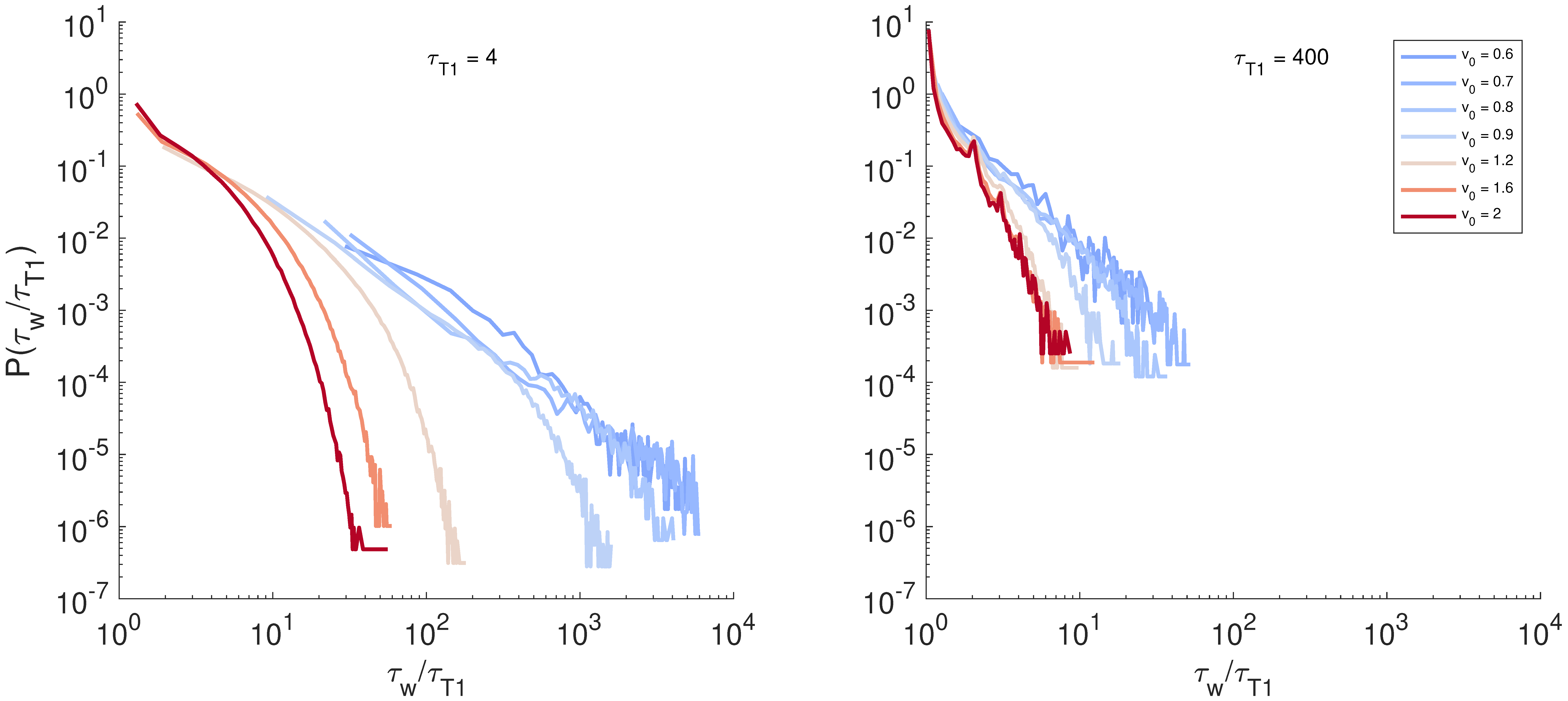}
\caption{
\textbf{Distributions of waiting times.} Here we show normalized probability distribution functions of measured waiting times \dti between successive T1 events involving individual cells at different \vz and two different \tti values. The distributions at small \tti and \vz near glass transition are much broader with heavy power law tails, while they are always quite narrow and without heavy tails for large \tti value.
}
\label{figS3}
\end{center}
\end{figure*}

\begin{figure*}[htbp]
\begin{center}
\includegraphics[width=1\columnwidth]{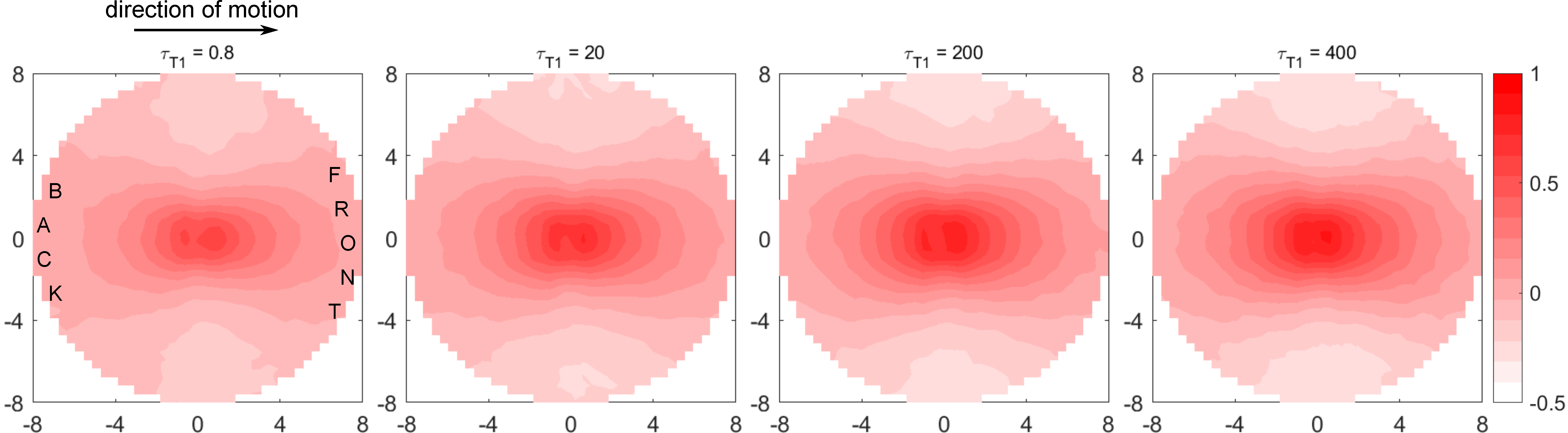}
\caption{
\textbf{Spatial directional correlation of cell motions around the fast cells in the tissue.} We follow a recent analysis~\cite{doi:10.1091/mbc.e16-05-0329}  that probes how the motion of any two cells are correlated as a function of both their distance and the direction from one cell, chosen as reference, to another. We find that this directional correlation, shown above for different \tti values at \vz$=1.6$, is low and nearly uniform in all directions, and decays very sharply within 1-2 cell diameters for low \tti. As \tti increases we see the correlation contours getting longer-ranged, polarized and elongated along the direction of motion of the reference cell. These features are classical signatures of predominant leader-follower behavior. For \tti$\geq 100$ we see very strong, anisotropic directional correlations that remain significant even at 6 cell diameters away from the reference cell. This approximate correlation length is consistent with that found from vector field analysis of cell streaming. These results show that the fast cells play major roles driving the cellular streaming observed for very large T1 delays.
}
\label{figS4}
\end{center}
\end{figure*}

\end{document}